\documentclass[twocolumn,10pt]{article}
\usepackage{usenix2019_v3}

\usepackage{tikz}
\usepackage{amsmath}

\usepackage{filecontents}
\usepackage{makecell}
\usepackage{multirow}
\usepackage{amssymb}
\usepackage{subfigure}
\usepackage{xspace}
\usepackage{microtype}
\microtypecontext{spacing=nonfrench}
\usepackage{hhline}
\usepackage{booktabs} 
\usepackage{xcolor}
\usepackage{algorithm}
\usepackage{comment}
\usepackage{algorithmic}
\usepackage{bm}
\usepackage{boxedminipage}

\definecolor{mauve}{rgb}{0.87, 0.63, 0.87}
\newif\ifcomments
\commentstrue

\definecolor{dkgreen}{rgb}{0,0.6,0}
\definecolor{gray}{rgb}{0.5,0.5,0.5}
\definecolor{orange}{rgb}{0.58,0.2,0}
\definecolor{mauve}{rgb}{0.58,0,0.82}
\definecolor{lightgray}{gray}{0.90}
\definecolor{ms-office-blue}{rgb}{0.36, 0.60, 0.83 } 

\newcommand{\authorcomment}[3]{\ifcomments\textcolor{#2}{#1 #3}\fi}

\newcommand{\modified}[1]{\authorcomment{}{black}{#1}}

\newcommand{\name}{ECO-LLM\xspace}
\newcommand{\fullname}{Edge-Cloud Orchestrator\xspace}
\newcommand{\emulator}{\textit{ECO-LLM Emulator}\xspace}
\newcommand{\runtime}{\textit{ECO-LLM Runtime}\xspace}

\newcommand{\ie}{\textit{i.e.},~}

\begin{document}

\date{}

\title{\Large \bf \name: Orchestration for Domain-specific Edge-Cloud Language Models}

\author{
{\bf Prasoon Patidar}$^{1}$ \quad {\bf Alex Crown}$^{2}$ \quad {\bf Kevin Hsieh}$^{2}$ \quad {\bf Yifei Xu}$^{2}$ \\
{\bf Tusher Chakraborty}$^{2}$ \quad {\bf Ranveer Chandra}$^{2}$ \quad {\bf Yuvraj Agarwal}$^{1}$ \\
\small{$^{1}$Carnegie Mellon University \quad $^{2}$Microsoft Research}
} 

\maketitle

\begin{abstract}
The remarkable performance of Large Language Models
(LLMs) has inspired many applications, which often necessitate edge-cloud collaboration due to connectivity, latency, and cost considerations. Traditional methods primarily focus on selecting the best LLM model for optimizing performance, while neglecting the critical interplay between the components of the LLM serving pipeline (context retrieval, query preprocessing, etc.) or the changing latency and cost constraints. We introduce \name~(\fullname for LLMs), a novel system that reframes this problem as a joint optimization challenge and solves it by systematically exploring component configurations and dynamically selecting optimal strategies at the query level.

\name consists of two components: (1) the \emulator, which efficiently explores the vast configuration space utilizing query clustering and pareto-optimal path selection, gathering domain-specific performance metrics without exhaustive evaluation; and (2) the \runtime, which leverages these metrics to dynamically select optimal resolution strategies for user queries while meeting user-defined Service Level Objectives (SLOs). 

\modified{We evaluate \name across five diverse domains on four hardware platforms. Compared to state-of-the-art model routing approaches, \name reduces costs by 60\% and latency by up to 6x while maintaining consistent performance (73-87\% accuracy) across domains, whereas model-routing approaches exhibit high variance (54-85\%)—degrading significantly on domains that require coordinated query preprocessing rather than model selection alone. ECO-LLM consistently meets user-defined SLOs while maintaining stable accuracy, and its stratified exploration achieves near-equivalent performance with up to 65\% fewer evaluations—making systematic domain adaptation practical.}

\end{abstract}

\section{Introduction}

The rapid advances in Large Language Models (LLMs) has led to the proliferation of AI assistants~\cite{future-of-consumer-edge-ai, apple-intelligence, copilot-pc, gemini-ai-assistant}. These AI assistants for smart homes, connected cars, and mobile devices benefit significantly from edge deployments. For instance, smart home assistants process voice commands locally for privacy, driving assistants require real-time data processing to navigate safely, minimizing reliance on cloud connectivity. Additionally, edge processing can reduce the cost of LLM inference on the cloud, a major concern for applications requiring frequent interactions with LLMs.

Despite the potential, developing an edge-AI assistant is challenging due to an ever expanding design space. For example, an end-to-end \emph{serving pipeline} can choose from a selection of LLM~\cite{phi-3, gemma, llama3}, retrieval-augmented generation (RAG)~\cite{rag} for fetching relevant context, and query/context refinement \cite{stepback,correctiverag}. Each of these components represents an active research area with rapid advances~\cite{rag-survey, model-compression-survey, llm-survey}. As we show, the selection of these components critically impacts overall performance and must consider factors such as edge device capabilities, domain-specific requirements, and user-defined constraints for cost and latency.

\modified{Current approaches to address this challenge primarily focus on optimizing individual components in isolation. Solutions such as RouteLLM \cite{routellm}, AutoMix \cite{automix} and HybridLLM \cite{hybridllm} have introduced model routing systems that optimize for selecting the best model for a given query, offering simple trade-offs such as quality versus cost. However, edge AI assistants inherently require domain-specific optimization, \ie automotive assistants require rapid responses with efficient content retrieval while smart home assistants must handle ambiguous queries requiring sophisticated reasoning. A critical insight missing from existing work is that optimal configurations require simultaneous selection across \emph{multiple pipeline} components, not just model selection. Some queries achieve higher accuracy with less context given to a powerful model, while others benefit from extensive retrieval with a smaller model. The challenge lies not just in finding these synergistic configurations, but in making domain-specific deployment efficient and systematic rather than requiring manual exploration for each new domain.}

\modified{To address this gap, we present \name, a deployment framework that transforms edge-cloud LLM optimizations from a manual, domain-by-domain effort into a systematic process. Rather than providing a pre-trained solution, ECO-LLM automatically explores query resolution paths for each domain and hardware setup, enabling per-query selection of optimal component configurations at runtime. Our key insight is that by treating component selection as a unified optimization problem and automating the exploration process, we can efficiently discover superior pareto frontiers balancing accuracy, cost, and latency for each specific domain.} 

\name consists of two complementary components: a \emulator and a \runtime. The \emulator provides a systematic exploration framework that efficiently characterizes different query resolution paths \modified{for a target domain and hardware setup} across multiple metrics. By intelligently sampling the vast configuration space without requiring exhaustive evaluation, it uncovers patterns that would remain hidden when optimizing components independently. Building on these insights, the \runtime dynamically selects optimal query resolution paths in real-time, combining domain-specific query encoding with constraint-aware path selection to consistently meet user-defined SLOs. \modified{This two-stage approach makes domain adaptation practical—the emulator handles the complex exploration once per domain/hardware combination, while the runtime efficiently serves queries using the discovered strategies.}

In developing \name, we addresses several key technical challenges. The system establishes a structured framework for representing and exploring the exponential configuration space of LLM serving pipelines, enabling comprehensive performance characterization without evaluating all possible component combinations. It introduces a novel approach for critical component identification that reduces the effective search space through impact-based analysis, making real-time decisions practical. Finally, it implements a robust path selection mechanism that generalizes performance patterns to unseen queries while strictly maintaining SLO guarantees. \modified{Our framework-based approach allows organizations to deploy edge AI assistants for new domains or hardware platforms efficiently, providing fine-grained control over latency and cost without the burden of manual optimization.}

We evaluate \name across four hardware platforms, Jetson Orin Nano, M1 Pro, M4, and RTX A4500 spanning resource-constrained to high-performance edge computing, and five diverse domains: automotive diagnostics, smart home automation, agriculture, technical support, and IoT security. Our results show that M4 class platforms (30-40 TOPS, 20-32 GB RAM) are a practical deployment target for responsive edge AI assistants—achieving sub-second to few-second latencies while the Jetson Orin Nano exhibits prohibitive response times (12+ seconds) and the RTX A4500 yields diminishing returns despite higher power/cooling needs. Across domains, \name reduces costs by 60\% and latency by up to 6x compared to state-of-the-art model routing approaches while maintaining comparable accuracy. Critically, \name delivers predictable performance (73-87\% accuracy) across all domains, whereas model-routing approaches show high variance (54-85\%)—degrading significantly on domains like smart home and technical support where queries benefit from coordinated preprocessing rather than model selection alone. Our ablation studies confirm that both \name's query understanding and adaptive path selection are necessary: removing the former degrades accuracy, while removing the latter forces suboptimal cost-latency tradeoffs. Finally, \name configurations achieve near-zero SLO violations when constraints are feasible, maintaining stable accuracy.

In summary, we make the following contributions:

\noindent 1. We formulate edge-cloud LLM deployment as a joint component selection problem, demonstrating that coordinated optimization across the serving pipeline delivers consistent performance where model-routing-only approaches exhibit high variance across domains.

\noindent 2. We built the \name Emulator, an exploration framework that efficiently characterizes query resolution paths through stratified budget allocation and critical component analysis, enabling systematic domain adaptation without exhaustive evaluation.

\noindent 3. We develop the \name Runtime, which combines domain-specific query encoding with constraint-aware path selection to dynamically select optimal resolution strategies while meeting user-defined SLOs.

\noindent 4. We evaluate \name across five domains and four hardware platforms, demonstrating 60\% cost reduction and up to 6× latency improvement over model-routing approaches, with predictable 73-87\% accuracy and consistent SLO attainment.

\vspace{-1em}

\section{Related Work}
\vspace{-1em}
The growing adoption of LLMs in edge applications has led to significant research addressing various aspects of LLM serving systems. We categorize existing approaches based on their primary focus and optimization strategies, and discuss how ECO-LLM relates to these approaches.

\begin{table}[t]
\centering
\footnotesize
\setlength{\tabcolsep}{2.5pt}
\caption{Comparing \name with different approaches for LLM inference. Unlike existing solutions that optimize individual components separately, \name addresses all of the requirements of an edge-cloud orchestration framework.}
\begin{tabular}{|l|c|c|c|c|c|}
\hline
\textbf{Approach} & \makecell{\textbf{Model}\\\textbf{Routing}} & \makecell{\textbf{Cost/}\\\textbf{Latency}} & \makecell{\textbf{Context/}\\\textbf{Query Opt.}} & \makecell{\textbf{Edge}\\\textbf{Deploy}} & \makecell{\textbf{Policy}\\\textbf{Constr.}} \\
\hline
\multicolumn{6}{|l|}{\textit{Model Routing Solutions}} \\
\hline
RouteLLM~\cite{routellm} & $\checkmark$ & $\checkmark$/$\times$ & $\times$/$\times$ & $\times$ & $\times$ \\
\hline
MetaLLM~\cite{metallm}\dag~ & $\checkmark$ & $\checkmark$/$\times$ & $\times$/$\times$ & $\times$ & $\times$ \\
\hline
LLMProxy~\cite{llmproxy}\dag~ & $\checkmark$ & $\checkmark$/$\times$ & $\times$/$\times$ & $\times$ & $\times$ \\
\hline
AutoMix~\cite{automix} & $\checkmark$ & $\checkmark$/$\times$ & $\times$/$\times$ & $\times$ & $\times$ \\
\hline
Eagle~\cite{eagle} & $\checkmark$ & $\checkmark$/$\checkmark$ & $\times$/$\times$ & $\times$ & $\times$ \\
\hline
TensorOpera~\cite{tensoropera} & $\checkmark$ & $\checkmark$/$\checkmark$ & $\times$/$\times$ & $\times$ & $\times$ \\
\hline
Cascade Router~\cite{cascade}\dag~ & $\checkmark$ & $\checkmark$/$\checkmark$ & $\times$/$\times$ & $\times$ & $\times$ \\
\hline
ScaleLLM~\cite{scalellm} & $\checkmark$ & $\times$/$\checkmark$ & $\times$/$\times$ & $\times$ & $\times$ \\
\hline
RouterDC~\cite{chen_routerdc_2024} & $\checkmark$ & $\times$/$\times$ & $\times$/$\times$ & $\times$ & $\times$ \\
\hline
EmbedLLM~\cite{zhuang_embedllm_2024} & $\checkmark$ & $\times$/$\times$ & $\times$/$\times$ & $\times$ & $\times$ \\
\hline
OptiRoute~\cite{piskala_dynamic_2024} & $\checkmark$ & $\checkmark$/$\checkmark$ & $\times$/$\times$ & $\times$ & $\checkmark$ \\
\hline
Routoo~\cite{mohammadshahi_routoo_2024}\dag~ & $\checkmark$ & $\checkmark$/$\times$ & $\times$/$\times$ & $\times$ & $\times$ \\
\hline
\multicolumn{6}{|l|}{\textit{Context Management Solutions}} \\
\hline
Teola~\cite{teola}\dag~ & $\times$ & $\times$/$\checkmark$ & $\checkmark$/$\checkmark$ & $\times$ & $\times$ \\
\hline
AutoRAG-HP~\cite{autorag} & $\times$ & $\checkmark$/$\times$ & $\checkmark$/$\checkmark$ & $\times$ & $\times$ \\
\hline
HyPA-RAG~\cite{hyparag} & $\times$ & $\checkmark$/$\times$ & $\checkmark$/$\checkmark$ & $\times$ & $\times$ \\
\hline
Mentor-KD~\cite{mentorkd} & $\checkmark$ & $\checkmark$/$\checkmark$ & $\checkmark$/$\checkmark$ & $\times$ & $\times$ \\
\hline
\multicolumn{6}{|l|}{\textit{Edge-Focused Solutions}} \\
\hline
HybridLLM~\cite{hybridllm} & $\checkmark$ & $\checkmark$/$\times$ & $\times$/$\times$ & $\checkmark$ & $\times$ \\
\hline
EdgeShard~\cite{edgeshard}\dag~ & $\checkmark$ & $\checkmark$/$\checkmark$ & $\times$/$\times$ & $\checkmark$ & $\times$ \\
\hline
PerLLM~\cite{yang_perllm_2024}\dag~ & $\checkmark$ & $\checkmark$/$\checkmark$ & $\times$/$\times$ & $\checkmark$ & $\checkmark$ \\
\hline
\textbf{\name} & \multirow{2}{*}{\bm{$\checkmark$}} & \multirow{2}{*}{\bm{$\checkmark$}/\bm{$\checkmark$}} & \multirow{2}{*}{\bm{$\checkmark$}/\bm{$\checkmark$}} & \multirow{2}{*}{\bm{$\checkmark$}} & \multirow{2}{*}{\bm{$\checkmark$}} \\[1.5pt]
\textbf{(Our Approach)} & & & & & \\
\hline
\end{tabular}
\label{tab:llm-comparison}
\vspace{-1em}
\end{table}

\textbf{Model Routing Systems:}
A significant body of work has focused on routing queries to appropriate language models based on query characteristics. Systems like RouteLLM~\cite{routellm} and AutoMix~\cite{automix} use routers trained on human preference datasets to direct queries to models with different capabilities, primarily optimizing cost-quality tradeoffs. Other approaches such as MetaLLM~\cite{metallm}, LLMProxy~\cite{llmproxy}, and Eagle~\cite{eagle} have introduced different strategies for model selection and scheduling to optimize inference performance.More recent approached like OptiRoute~\cite{piskala_dynamic_2024}  route tasks to the optimal LLM based on detailed user-defined requirements. These systems effectively address model selection as a standalone optimization problem. However, they don't explore how model selection interacts with other aspects such as context retrieval or query processing. Additionally, most of these approaches focus on cloud-based deployment, with less attention to edge environments.

\textbf{Context Management and Query Optimization:}
Another research direction has explored optimizing context management and query processing in LLM systems. Frameworks like Teola~\cite{teola} introduce orchestration systems that optimize dataflow representations of query workflows, while approaches like AutoRAG-HP~\cite{autorag} and HyPA-RAG~\cite{hyparag} focus on hyper-parameter tuning for retrieval-augmented generation. Mentor-KD~\cite{mentorkd} takes a different approach by enhancing smaller models through knowledge distillation. These solutions provide valuable insights into optimizing specific components of the pipeline but generally consider these optimizations separately from model selection and edge deployment considerations.

\textbf{Edge-Focused Deployments:}
Recent works have begun addressing the challenges of edge deployment for LLMs. HybridLLM~\cite{hybridllm} proposes a cost-efficient routing approach for edge environments.  EdgeShard~\cite{edgeshard}\dag~ introduces a partitioning strategy that distributes model components between edge and cloud, while PerLLM~\cite{yang_perllm_2024}\dag~ utilizes a multi-arm bandit optimization to schedule inference tasks across edge and cloud. More recent advances in edge-cloud collaborative frameworks include Edge-LLM~\cite{yu_edge-llm_2024}, EdgeFM~\cite{yang_edgefm_2023}, and CE-CoLLM~\cite{jin_ce-collm_2025}\dag~, which explore sophisticated approaches ranging from efficient model adaptation through layerwise compression, using a foundation model for open-set recognition, and enabling collaborative edge-cloud inference strategies. While these approaches make important contributions to edge-specific deployment challenges, they focus primarily on model distribution, and collaborative inference strategies. They generally give less attention to the holistic optimization of context retrieval and query manipulation components alongside model selection, which can be critical for maintaining response quality in resource-constrained environments. These model-centric optimizations can be effectively integrated with \name to provide domain-specific pipeline optimization, complementing their focus on efficient model execution with systematic component selection.

\footnotetext[1]{\dag~ arXiv technical reports}
\textbf{Research Gap and Our Contributions:}

As Table \ref{tab:llm-comparison} illustrates, existing approaches reflect a gap in the research landscape—the lack of integrative frameworks that address the interplay between these components. While each component optimization is valuable individually, their interactions remain underexplored. For instance, how context retrieval strategies should adapt to different model choices, or how query processing needs change based on edge deployment conditions. \name's contribution lies in connecting these disparate optimization domains into a unified framework. By integrating policy-driven constraints with component-level optimizations, our work builds upon these works. This approach allows specialized systems to be composed in ways that respect their interdependencies, potentially offering a more complete perspective on LLM serving optimization for domain-specific deployments.

\vspace{-1em}

\section{System Design}
\vspace{-1em}

We envision a framework capable of addressing the diverse challenges and requirements of Edge AI assistants across various domains. This framework will integrate local processing (edge databases, on-device inference, caching) with cloud resources (larger language models, extensive databases).
The "Edge" refers to the computing platforms one step removed from the most constrained environments (microcontrollers, basic sensors, etc.), i.e., devices that can run Small Language Models (SLMs) and host local databases, including consumer laptops with NPUs, automotive computing units, smart home hubs, and industrial edge servers—while maintaining the benefits of local processing.

Each domain presents unique hurdles: automotive assistants demand rapid, precise responses regarding vehicle systems, emphasizing efficient content retrieval, whereas smart home assistants grapple with ambiguous queries necessitating more sophisticated reasoning. Furthermore, requirements can vary even within a single domain, depending on the query. For instance, an automotive assistant in a smart car benefits from web data retrieval for weather updates and navigation, but relies on local, specialized databases for vehicle-specific queries.

\subsection{Problem Formulation}

\begin{figure}[t]
\begin{center}
    \includegraphics[width=0.47\textwidth]{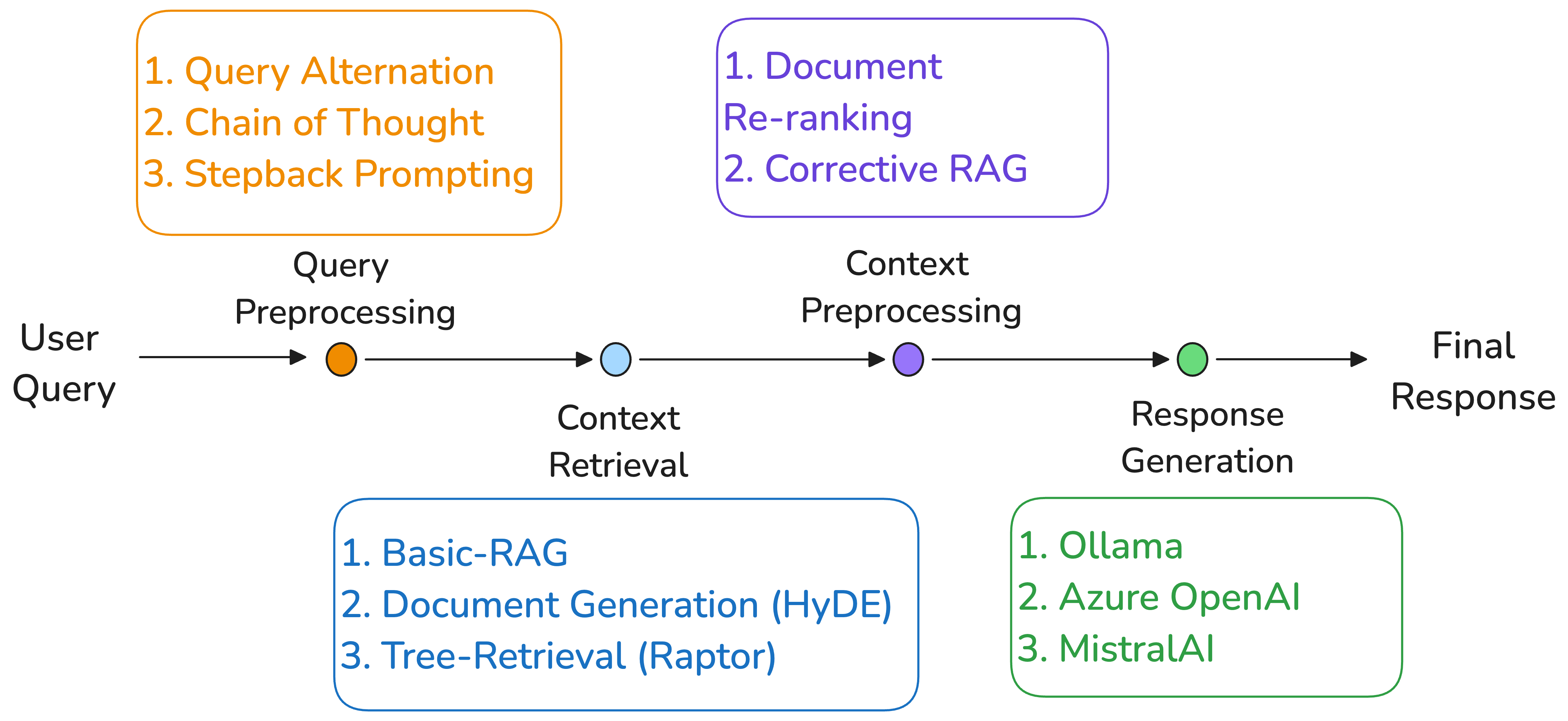}
\end{center}
\caption{Sequence of components (or \textit{query resolution path}) to resolve a user query with edge AI assistants.}
\label{fig:query_resolution_path}
\vspace{-1.5em}
\end{figure}

We define a query resolution path $P$ as the sequence of operations that transforms a user query into a final response (See Figure \ref{fig:query_resolution_path}). Each path is an ordered tuple $P = (M_q, M_r, M_c, M_m)$, where $M_q$ represents query processing, $M_r$ represents retrieval strategy, $M_c$ represents context processing, and $M_m$ represents model selection. Each module $M_i$ has multiple potential implementations $I_i$ and parameter configurations $C_{i,j}$ for each implementation $j \in I_i$. A concrete path instantiation involves selecting both implementations and configurations: $P = ((q, \theta_q), (r, \theta_r), (c, \theta_c), (m, \theta_m))$.

The space of all possible paths $\mathcal{P}$ grows combinatorially with the number of modules, implementations, and configurations (see Table \ref{tab:configuration_examples}):
\begin{equation}
|\mathcal{P}| = \prod_{i \in \{q,r,c,m\}} \sum_{j \in I_i} |C_{i,j}|
\label{eq:configuration_space}
\end{equation}

For each query $q \in Q$ and path $P \in \mathcal{P}$, we define three key performance metrics:
\begin{equation}
\begin{aligned}
\text{Accuracy}(q, P) &: Q \times \mathcal{P} \rightarrow [0, 1] \\
\text{Latency}(q, P) &: Q \times \mathcal{P} \rightarrow \mathbb{R}^+ \\
\text{Cost}(q, P) &: Q \times \mathcal{P} \rightarrow \mathbb{R}^+
\end{aligned}
\end{equation}

\modified{We measure latency as Time to First Token (TTFT), which denotes the time until the user receives the first response token. TTFT includes query preprocessing, context retrieval and postprocessing, model inference, and \name's path selection overhead. We estimate cost based on token usage:}
\begin{equation}
\mathcal{C}(q, P) = \alpha_P \cdot |q| + \beta_P \cdot \text{max\_tokens}
\end{equation}
\modified{where $\alpha_P$ and $\beta_P$ are model-specific pricing for input and output tokens.}

Service Level Objectives (SLOs) impose constraints on latency and cost:
\begin{equation}
\begin{aligned}
\text{Latency}(q, P) &\leq \mathcal{L}_{\text{max}} \\
\text{Cost}(q, P) &\leq \mathcal{C}_{\text{max}}
\end{aligned}
\end{equation}

We identify two challenges in edge-cloud orchestration:

1) Given a domain-specific dataset $D$ of representative queries, we must efficiently explore the performance characteristics of different paths within a given budget $B$.
\begin{equation}
\begin{aligned}
\max_{P_{subset} \subset \mathcal{P}} &\quad \text{InformationGain}(P_{subset}) \\
\text{s.t.} &\quad |P_{subset}| \leq B
\end{aligned}
\end{equation}

2) For a new, previously unseen query $q_{new}$, we need to select a path that satisfies SLO constraints while maximizing accuracy in real time:
\begin{equation}
\begin{aligned}
\max_{P \in \mathcal{P}} &\quad \text{Accuracy}(q_{new}, P) \\
\text{s.t.} &\quad \text{Latency}(q_{new}, P) \leq \mathcal{L}_{\text{max}} \\
&\quad \text{Cost}(q_{new}, P) \leq \mathcal{C}_{\text{max}}
\end{aligned}
\label{eq:optimization_fun}
\end{equation}

\begin{figure}[t]
\begin{center}
 \subfigure[Smart Home Assistant]{
    \centering
    \label{fig:smarthome_path_analysis}
    \includegraphics[width=0.22\textwidth]{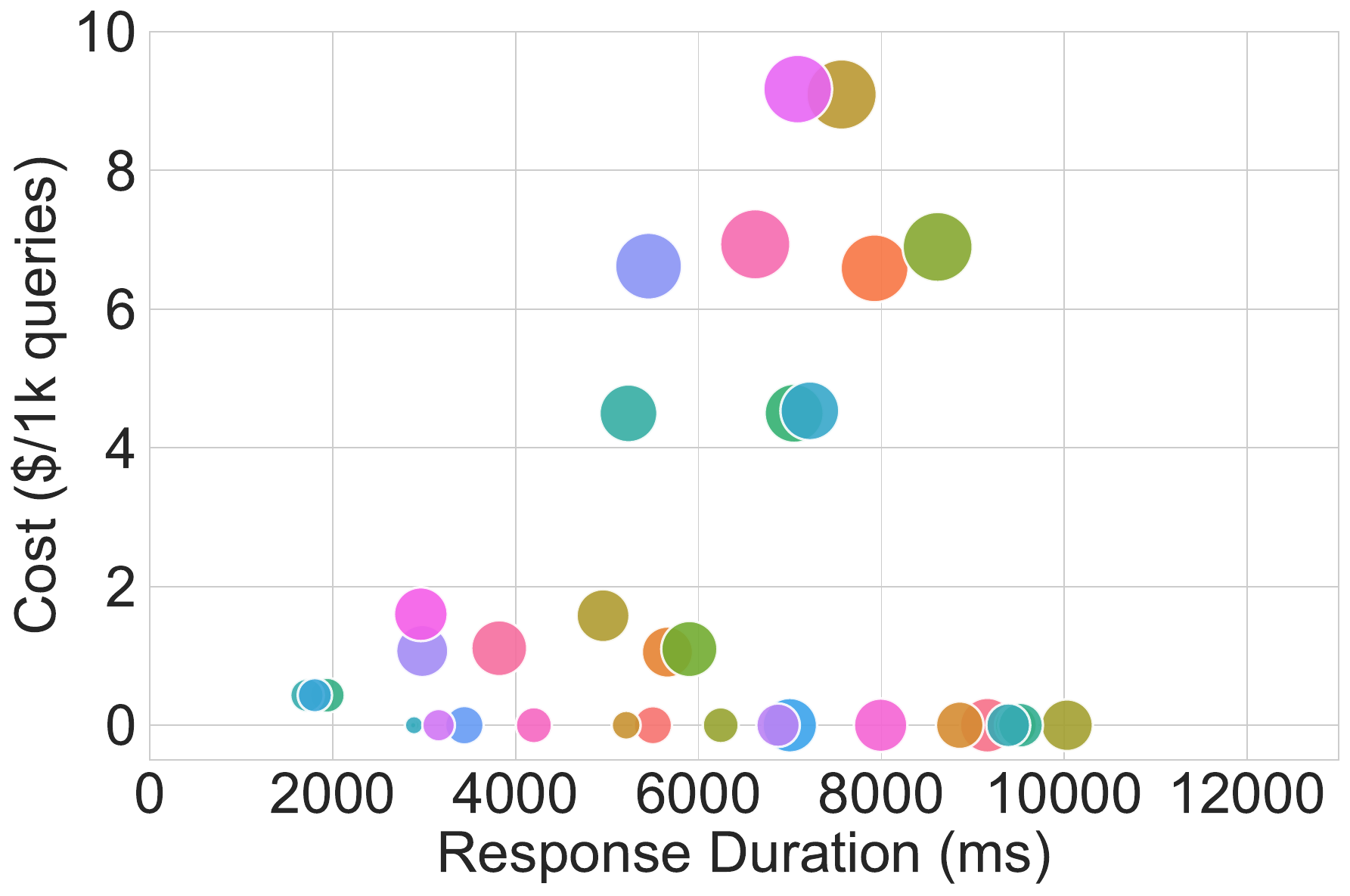}}
 \hfill
 \subfigure[Automotive Assistant]{
     \centering
     \label{fig:smartcar_path_analysis}
     \includegraphics[width=0.22\textwidth]{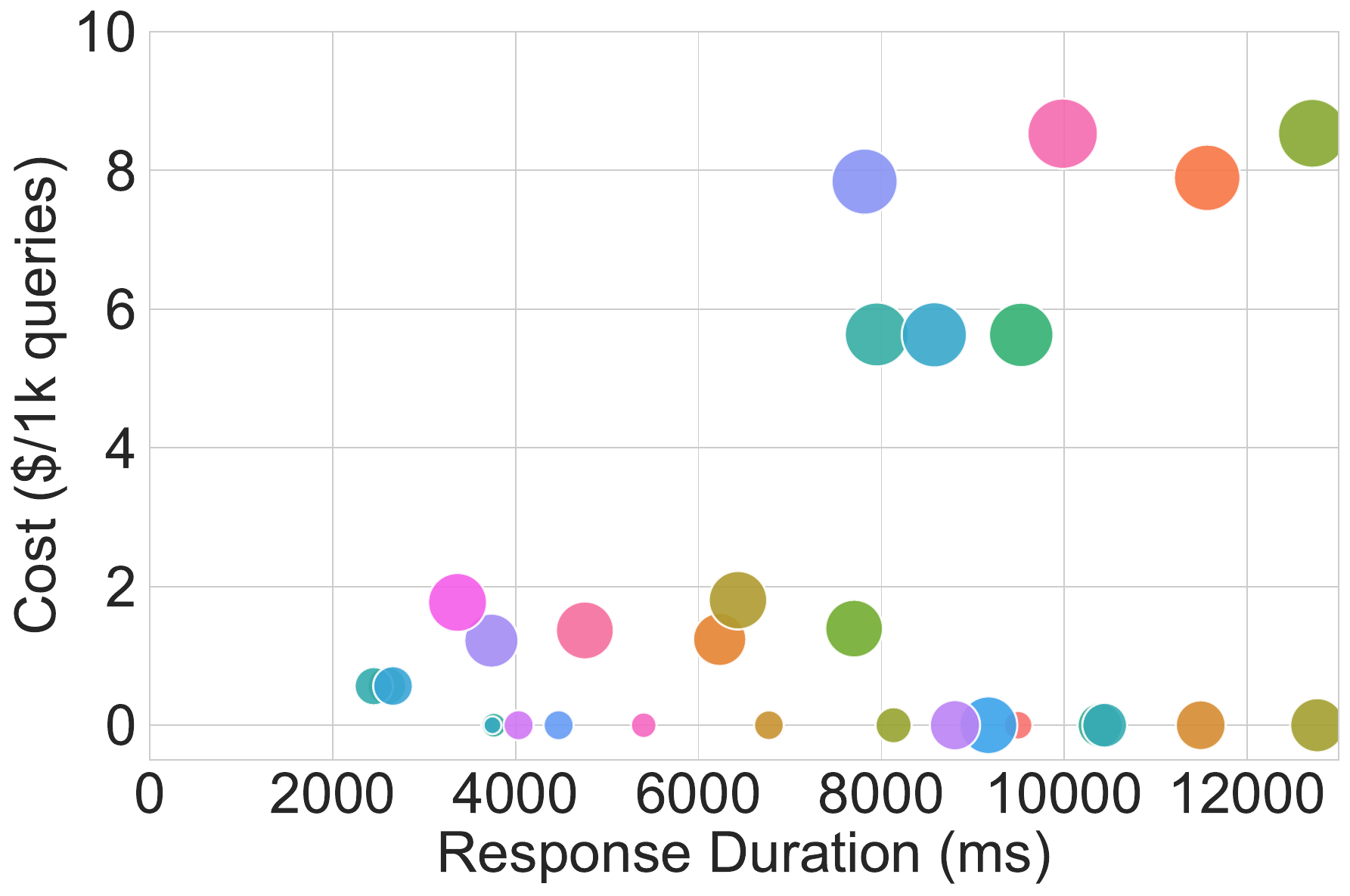}}
\end{center}
\caption{Performance variations in query resolution paths for smart home and Automotive AI assistants due to differences in domain-specific queries. Each color represents a single path (or configuration), and the marker size represents the quality of response. Different paths have optimal response quality for different domains.}
\label{fig:path_analysis}
\end{figure}

A key insight is that the effectiveness of resolution paths varies significantly across domains (see Figure \ref{fig:path_analysis}). The same path yields different latency-cost-accuracy tradeoffs for smart homes versus automotive assistants. Moreover, even within a single domain, optimal paths vary significantly by query characteristics. While simple heuristics—such as routing complex queries to cloud models or always using maximum context retrieval—can achieve reasonable accuracy, they could incur much higher costs and are unsuitable for scenarios with strict latency requirements. Finally, the optimization challenge is compounded by component interdependencies where model performance depends critically on query preprocessing and retrieved context.


\begin{figure*}[t]
  \centering
  \hfill
  \includegraphics[width=0.95\textwidth]{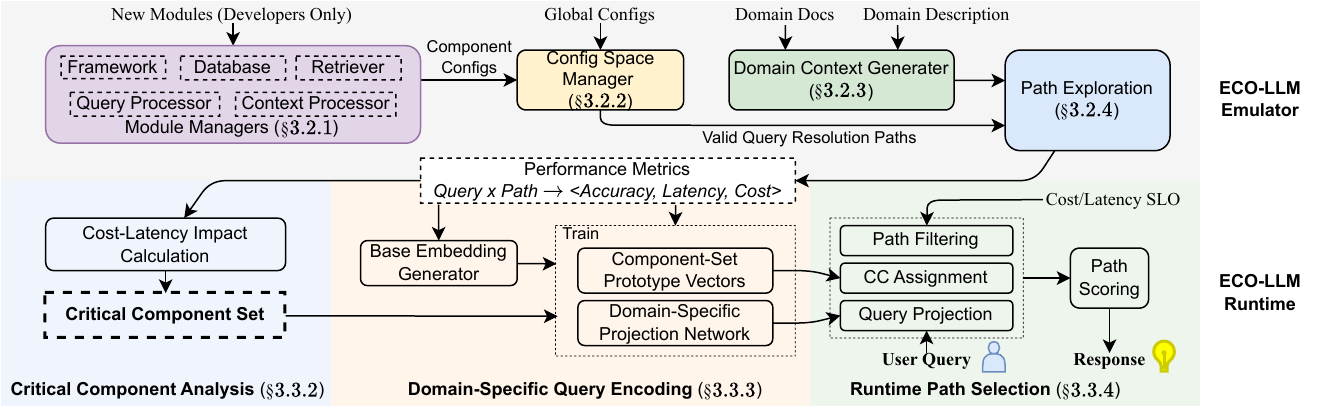}
  \hfill
  \vspace{-0.18in}
  \caption{Overall Architecture of \name, showing the components of the \emulator and the \runtime.}
  \vspace{-0.09in}
  \label{fig:overview}
\end{figure*}

\begin{table*}[htbp]
\centering
\footnotesize
\caption{Examples of implementations ($j \in I_i$) for each module $M_i$ and their configurable parameters. Parameters can be defined as static values, parameter ranges to sweep through during exploration, or dynamically resolved at runtime via methods (e.g., available models). Each query resolution path $P = ((q, \theta_q), (r, \theta_r), (c, \theta_c), (m, \theta_m))$ represents a specific implementation and parameter configuration choice. The combinatorial explosion of options leads to an exponential growth in the configuration space, making systematic exploration critical.}
\begin{tabular}{|p{3.2cm}|p{4cm}|p{8cm}|}
\hline
\textbf{Module Type} & \textbf{Implementations} & \textbf{Configurable Options} \\
\hline
\multirow{2}{*}{\parbox{3.2cm}{Model Frameworks}} 
& Ollama~\cite{ollama} & model name, temperature, context length \\
\cline{2-3}
& Azure OpenAI~\cite{azureopenai} & api key, deployment name, temperature, max tokens, presence penalty \\
\cline{2-3}
\hline
\multirow{2}{*}{\parbox{3.2cm}{Query Processing\\ (Pre-retrieval)}} 
& Chain of Thought~\cite{cot} & reasoning depth, intermediate steps, verification strategy \\
\cline{2-3}
& Step-back Prompting~\cite{stepback} & abstraction level, context breadth \\
\hline
\multirow{2}{*}{\parbox{3.2cm}{Context Retrieval}} 
& Basic RAG~\cite{basicrag} & chunk size, overlap ratio, top-k \\
\cline{2-3}
& HyDE~\cite{hyde} & hypothesis model, hypothesis count, aggregation strategy \\
\cline{2-3}
\hline
\multirow{2}{*}{\parbox{3.2cm}{Context Processing \\ (Post-retrieval)}} 
& Corrective RAG~\cite{correctiverag} & correction threshold, iteration limit, confidence score \\
\cline{2-3}
& Document Reranking~\cite{reranking} & reranking model, score threshold, cross-attention weight \\
\hline
\end{tabular}
\label{tab:configuration_examples}
\end{table*}

\subsection{\emulator}

The \emulator addresses the challenge of efficiently exploring the exponential space of query resolution paths to identify optimal configurations for domain-specific queries. Instead of implementing the components themselves, \emulator provides a framework for systematically evaluating different combinations of configurations.

\subsubsection{Modular Architecture}

Given that new language models, retrieval techniques, and processing methods emerge monthly, we designed \emulator to be extensible. The architecture consists of five standardized interfaces implemented as module managers: Framework Manager ($M_m$) for language model interactions, Database Manager for vector storage, Query Processor Manager ($M_q$) for query transformations, Retriever Manager ($M_r$) for context extraction, and Context Processor Manager ($M_c$) for post-retrieval refinement. 
This design offers three key advantages: First, it enables easy integration of emerging technologies, i.e., new retrieval methods or language models can be added by implementing the appropriate interface without modifying other system components. Second, it facilitates systematic experimentation by allowing different combinations of implementations to be evaluated under consistent conditions. Third, it provides abstraction across deployment environments, with each manager supporting both edge and cloud implementations through the same interface. The modules interact through well-defined boundaries while leveraging each other's capabilities—for example, retrievers can use language models via the Framework Manager to enhance search effectiveness. This design ensures that \emulator can continuously incorporate state-of-the-art components while maintaining systematic evaluation capabilities.

\subsubsection{Configuration Space Management}

\emulator employs a hierarchical configuration system with component-level and global-level specifications to keep configuration management tractable. At the component level, each implementation for a module specifies three types of parameters: static values for standard operation, parameter ranges to sweep through during experimentation, and dynamic values resolved at runtime. For example, a language model framework might include static temperature values, model context lengths to sweep through, and dynamically resolved model names based on what is available at runtime. At the global level, configurations define which implementations to include for each module, execution settings, and evaluation strategies. The global configuration can also specify null options to allow the system to skip certain operations entirely.

\subsubsection{Domain Context Generation}

\modified{Many domains such as automotive diagnostics or interactions with industrial equipment and medical devices have technical documentation but lack query datasets for evaluating LLM pipelines. Prior approaches propose creating these query datasets manually by domain experts which is time consuming and tedious. \emulator addresses this query dataset creation through its \textit{Context Generator}, which creates training datasets from domain documentation (e.g., vehicle manuals) and a brief domain description.}

\modified{The Context Generator uses LLMs to produce questions from documentation through content extraction, chunking, and prompt-guided query generation. It creates six query types targeting different information needs: retrieval questions for specific facts, explanation questions for causal relationships, analysis questions for multi-factor reasoning, solving questions for troubleshooting procedures, comparison questions for evaluating alternatives, and recommendation questions for optimizing under constraints.} 

\modified{For example, retrieval questions target specific information (\textit{``What is the implication of unauthorized modifications on operating authority?''}), while analysis questions require reasoning across multiple factors (\textit{``What are safety implications if the Reverse Brake Assist warning persists despite troubleshooting?''}). Recommendation questions consider multiple constraints (\textit{``How should I schedule charging to minimize costs while ensuring morning readiness?''}). These queries combine multiple system aspects rather than following simple templates. Each query $Q_i$ includes metadata $(T_i, C_i, E_i)$ representing query type, expected context/reference answer, and evaluation guidelines for automated response quality assessment. Application owners may optionally review generated answers to further improve quality. This approach requires only documentation that organizations deploying domain-specific AI assistants typically already possess.}





\subsubsection{Path Exploration}

\modified{\emulator characterizes query resolution paths across the configuration space defined in Equation \ref{eq:configuration_space}. Exhaustive evaluation of all query-path combinations becomes computationally prohibitive for realistic deployments.
To address this, \emulator uses adaptive Stratified Budget Allocation (SBA), outlined in Algorithm \ref{alg:sba}. The approach operates in two stages: first, it selects representative queries from each query type and evaluates all paths on these queries to identify high-performing paths per type; second, for remaining queries, it evaluates only the most promising paths identified in the initial stage. The budget factor $B$ provides unified control over this two-stage exploration, independently scaling query selection for breadth ($B \times \sqrt{|Q|}$ queries) and path selection for depth ($B \times \sqrt{|\mathcal{P}|}$ paths per query). This sublinear scaling reduces total evaluations from $O(|Q| \times |\mathcal{P}|)$ to $O(\sqrt{|Q|} \times |\mathcal{P}| + |Q| \times \sqrt{|\mathcal{P}|})$, enabling exploration within computational constraints.}

\modified{Representative queries are selected using k-means clustering on sentence transformer embeddings, choosing queries closest to each cluster centroid to ensure semantic diversity within each query type. Path ranking uses accuracy as the primary criterion, with cost or latency as secondary tiebreaker depending on optimization strategy ($\lambda$ in Section \ref{subsubsec:cca}).}

\begin{algorithm}[t]
\caption{Adaptive Stratified Budget Allocation}
\label{alg:sba}
\begin{algorithmic}[1]
\REQUIRE Queries $Q$, Paths $\mathcal{P}$, Budget factor $B$
\ENSURE Evaluation results $R$
\STATE $Q_{rep} \gets \text{StratifiedSample}(Q, B)$ \COMMENT{k-means on embeddings by type}
\STATE Evaluate all paths $\mathcal{P}$ for queries $Q_{rep}$
\STATE Group and rank paths by accuracy, then cost/latency per query type
\STATE $k \gets \max(1, B \cdot \sqrt{|\mathcal{P}|})$ \COMMENT{Paths per remaining query}
\FOR{$q \in Q \setminus Q_{rep}$}
    \STATE $P_{select} \gets$ $k$ paths from top-ranked for $q$'s type + random samples
    \STATE Evaluate only paths in $P_{select}$ for query $q$
\ENDFOR
\end{algorithmic}
\end{algorithm}

\emulator implements prefix caching to reduce redundant computation. When two paths differ only in their final component (e.g., both use step-back prompting and HyDE retrieval but different models), the system caches intermediate results and only executes the component that differs. This reduces computation by 30-50\% in domains with expensive retrieval operations. For each evaluated path, \emulator collects performance metrics (accuracy, latency, cost) for use by the \runtime.

\subsection{\runtime}
\label{sec:orchestration_design}

\subsubsection{Problem Formulation}
\modified{The \runtime addresses path selection from Section 3.1 (Equation \ref{eq:optimization_fun}): for each unseen query, select a path maximizing accuracy while meeting latency and cost SLO constraints. The key challenge is that accuracy $\mathcal{A}(q,P)$ is unknown for new queries and the solution space grows exponentially with components. The \runtime solves this through three stages: Critical Component Analysis (CCA) identifies which components critically affect accuracy for each query type, Domain-Specific Query Encoding (DSQE) learns to predict component requirements for unseen queries, and Runtime Path Selection (RPS) chooses optimal paths meeting SLO constraints in real-time.}

\subsubsection{Critical Component Analysis (CCA)}
\label{subsubsec:cca}

The CCA module identifies which subset of components are essential for individual queries through performance evaluation and ablation studies. The key insight is that for each query, only a subset of components critically impacts response quality.

\begin{algorithm}[t]
\caption{Critical Component Analysis (CCA)}
\label{alg:cca}
\begin{algorithmic}[1]
\REQUIRE Query set $Q$, Component options $\mathcal{O} = \{M_q, M_r, M_c, M_m\}$, Accuracy threshold $\tau$, Latency preference $\lambda$
\ENSURE Critical component mapping $\Phi: Q \rightarrow 2^{\mathcal{O}}$
\STATE $\Phi \gets \emptyset$
\FOR{each $q \in Q$}
    \STATE $P^* \gets \text{FindBestPath}(q, \mathcal{O}, \lambda)$ \COMMENT{Find path with highest accuracy}
    \STATE $criticalComps \gets \emptyset$
    \FOR{each component type $t \in \{q, r, c, m\}$}
        \STATE $v \gets P^*[t]$ \COMMENT{Component value in best path}
        \STATE $impactScore \gets \text{CalculateImpact}(q, t, v, \mathcal{O}, P^*)$
        \IF{$impactScore \geq \tau$}
            \STATE $criticalComps \gets criticalComps \cup \{(t, v)\}$
        \ENDIF
    \ENDFOR
    \STATE $\Phi[q] \gets criticalComps$
\ENDFOR
\RETURN $\Phi$
\end{algorithmic}
\end{algorithm}

Algorithm \ref{alg:cca} outlines the CCA process. For each query (lines 2-12), we first identify the best-performing path $P^*$ (line 3) using a lexicographic optimization approach that prioritizes accuracy first, then optimizes for either latency or cost among paths (controlled by $\lambda \in \{0,1\}$ with comparable accuracy (within 1\% tolerance). When $\lambda = 1$, the system optimizes for latency; when $\lambda = 0$, it optimizes for cost. 

For each component type (lines 5-10), we calculate its impact on response quality:

\begin{equation}
\text{Impact}(q, t, v) = \mathcal{A}_{with}(q, t, v) - \mathcal{A}_{without}(q, t, v)
\end{equation}

where $\mathcal{A}_{with}$ and $\mathcal{A}_{without}$ represent the mean accuracy with component value $v$ fixed versus using alternatives:

\begin{equation}
\begin{aligned}
\mathcal{A}_{with}(q, t, v) = \frac{1}{|\mathcal{P}_{t=v}|} \sum_{P \in \mathcal{P}_{t=v}} \mathcal{A}(q, P)
\end{aligned}
\end{equation}
\begin{equation}
\begin{aligned}
\mathcal{A}_{without}(q, t, v) = \frac{1}{|\mathcal{P}_{t \neq v}|} \sum_{P \in \mathcal{P}_{t \neq v}} \mathcal{A}(q, P)
\end{aligned}
\end{equation}

Subset of components with impact scores exceeding threshold $\tau$ are marked as critical (lines 8-9). The result is a mapping $\Phi$ from queries to their critical components (line 11), which is used to train our domain-specific query encoder.

\subsubsection{Domain-Specific Query Encoding (DSQE)}

\modified{CCA identifies $K$ distinct critical component sets across training queries. Each set specifies which implementations and configurations from the path tuple (Section 3.1) critically affect query accuracy—for instance, one set might be \{(HyDE, $\theta_r$), (LLMLingua, $\theta_c$), (gpt-4o, $\theta_m$)\} while another is \{(BasicRAG, $\theta_r'$), (null, $\theta_c$), (llama-8b, $\theta_m'$)\}. DSQE learns to predict which of these $K$ sets a new query requires. The challenge is that semantic similarity does not reliably indicate component requirements: queries like ``Turn off the bedroom lights'' and ``Why won't bedroom lights turn off?'' are semantically similar but require different components. DSQE addresses this by training prototype vectors representing each component set and learning a projection that maps queries to their appropriate prototypes based on domain-specific patterns rather than semantic similarity alone. DSQE comprises three components:}

\textbf{1. Base Embedding Generator:} We use SentenceTransformer \cite{sentencetransformer} to generate dense vector representations $e_q \in \mathbb{R}^d$ for each query $q$, capturing semantic content as a start. 

\textbf{2. Projection Network:} We train a multilayer perceptron $f_\theta: \mathbb{R}^d \rightarrow \mathbb{R}^d$ that transforms semantic embeddings into a specialized space where queries requiring similar component sets cluster together:
\begin{equation}
f_\theta(e_q) = f_n(\ldots f_2(f_1(e_q)))
\end{equation}
where each layer $f_i$ applies:
\begin{equation}
f_i(x) = \text{ReLU}(\text{Dropout}(W_i x + b_i))
\end{equation}
with learnable parameters $\theta = \{W_i, b_i\}_{i=1}^n$. This projection learns domain-specific patterns that map queries to component requirements.

\textbf{3. Component-Set Prototypes:} We learn $K$ prototype vectors $\{v_k\}_{k=1}^K$ where each $v_k \in \mathbb{R}^d$ represents one of the distinct critical component sets identified by CCA. These prototypes are learnable parameters optimized jointly with the projection network.

We optimize the projection network and prototype vectors using three complementary objectives:
\begin{equation}
\mathcal{L}_{total} = \mathcal{L}_{contrast} + \alpha \cdot \mathcal{L}_{diversity} + \beta \cdot \mathcal{L}_{reg}
\end{equation}
The contrastive loss encourages queries to map close to their corresponding component set's prototype. The diversity loss promotes separation between different prototypes, preventing collapse into similar representations. The regularization term prevents overfitting. During training, each query is labeled with its critical component set from CCA, and the projection network learns to encode queries such that their similarity to prototypes indicates component requirements. For inference, the system projects the query embedding, computes similarities to all $K$ prototypes, and selects the component set corresponding to the most similar prototype.

\subsubsection{Runtime Path Selection (Online)}

The RPS module leverages the trained DSQE to make real-time path selection decisions for incoming queries, balancing response quality with SLO constraints, as outlined in Algorithm \ref{alg:path_selection}.

\begin{algorithm}[t]
\caption{Runtime Path Selection (RPS)}
\label{alg:path_selection}
\begin{algorithmic}[1]
\REQUIRE New query $q_{new}$, Component options $\mathcal{O}$, Trained DSQE $f_\theta$, Prototype vectors $\{v_k\}$, Training data $D$, SLO constraints $(\mathcal{L}_{max}, \mathcal{C}_{max})$
\ENSURE Selected path $P_{selected}$
\STATE $e_{new} \gets f_\theta(e_{q_{new}})$ \COMMENT{Project query embedding}
\STATE $k^* \gets \underset{k}{\text{argmax}} \, \text{sim}(e_{new}, v_k)$ \COMMENT{Find nearest prototype}
\STATE $criticalComps \gets \text{ComponentSet}(k^*)$
\STATE $validPaths \gets \emptyset$
\FOR{each $P \in \mathcal{P}$}
    \IF{$\mathcal{L}(P) \leq \mathcal{L}_{max}$ \AND $\mathcal{C}(P) \leq \mathcal{C}_{max}$ \AND $criticalComps \subseteq P$}
        \STATE $validPaths \gets validPaths \cup \{P\}$
    \ENDIF
\ENDFOR
\IF{$validPaths = \emptyset$}
    \RETURN $\text{DefaultPath}(criticalComps, \mathcal{O}, D, \mathcal{L}_{max}, \mathcal{C}_{max})$
\ENDIF
\FOR{each $P \in validPaths$}
    \STATE $pathScores[P] \gets \text{ScorePath}(P, e_{new}, D)$
\ENDFOR
\RETURN $\underset{P \in validPaths}{\text{argmax}} \, pathScores[P]$
\end{algorithmic}
\end{algorithm}

Algorithm \ref{alg:path_selection} comprises three key steps. First, the system projects the new query embedding using the trained DSQE model and identifies the most similar prototype (lines 1--3), immediately revealing critical component requirements based on patterns from training queries. Second, path filtering (lines 4--9) generates valid paths that meet both SLO constraints and critical component requirements:

\begin{equation}
\begin{aligned}
\mathcal{P}_{valid} = \{P \in \mathcal{P} \mid & \mathcal{L}(P) \leq \mathcal{L}_{max} \land \\
& \mathcal{C}(P) \leq \mathcal{C}_{max} \land \\
& criticalComps \subseteq P\}
\end{aligned}
\end{equation}

Third, path scoring (lines 12--14) uses nearest neighbor analysis on training queries:
\begin{equation}
\text{Score}(P, q_{new}) = \sum_{q \in NN_k(q_{new})} w_q \cdot \mathbb{I}[P_q = P] \cdot \mathcal{A}(q, P_q)
\end{equation}
where $NN_k(q_{new})$ returns $k$ nearest training queries, $w_q$ is the similarity weight, $\mathbb{I}$ is the indicator function, and $P_q$ is the best path for query $q$. The system selects the path with the highest score (line 15).

For out-of-distribution queries with no valid paths (lines 10--11), a fallback mechanism selects components based on global performance statistics while respecting critical components, choosing options that exceed an accuracy threshold $\tau_{acc}$ while minimizing cost.



\vspace{-1em}

\section{Implementation}
\label{sec:implementation}
\vspace{-1em}

We implemented both \emulator and \runtime as frameworks designed to simplify integration into development workflows. Our implementation focuses on maintainability, extensibility, and practical deployment. We plan to open-source the code upon publication.

We developed \emulator, a 5k-line command-line application in Python, for experiment execution, context generation, and path evaluation. 
We implemented a hierarchical configuration system that separates module-specific configurations from global experiment parameters. Each emulation configuration receives a unique build identifier to track associated parameters, results, and artifacts. This design supports multiple parallel configurations for different scenarios or environments, and asynchronous execution enables parallel evaluation across compute resources.

We implemented key components across all module managers. For query processing, we use stepback prompting \cite{stepback} and LLMLingua \cite{llmlingua} for model-based query compression. Our context retrieval includes basic RAG \cite{basicrag} and HyDE \cite{hyde}. For context processing, we use model-based compression \cite{llmlingua} and Corrective-RAG \cite{correctiverag} to improve accuracy. We integrate Ollama \cite{ollama} for edge deployment and OpenAI \cite{openai} for cloud processing. We extended DeepEval \cite{deepeval} for LLM evaluation (similar to G-Eval metrics \cite{geval}), employing an ensemble of two LLM judges (GPT-4o and Gemini-2.5-Flash) to compare generated responses against ground truth answers from our datasets. To mitigate self-preference bias \cite{wataoka2025selfpreferencebias}, neither judge model appears as a cloud option in our inference pipeline.

We implemented \runtime as an OpenAI compatible server (3k lines of code), specifically designed for edge deployment. Our implementation extends the standard OpenAI API specification
by adding support for build identifiers, SLO specification parameters, system state reporting capabilities, and enhanced error logging for deployment debugging.
Our implementation provides developers with the tools and documentation to integrate new components and target specific edge deployment requirements, while maintaining flexibility to handle diverse deployment scenarios and usage patterns.

\vspace{-1em}

\section{Evaluation}
\label{sec:evaluation}
\vspace{-1em}

Our evaluation aims to answer four research questions: (1) How does \name perform across diverse edge hardware platforms? (2) Can \name generalize across diverse application domains? (3) Which components of \name are important for its overall performance? 
(4) Can \name maintain SLOs under cost and latency constraints?

\subsection{Experiment Setup}
\textbf{Datasets.} We evaluate \name on five diverse domains. For \textbf{automotive assistants}, we use 3,000 pages of technical content from seven manufacturers covering diagnostics, maintenance, and troubleshooting. For \textbf{smarthome assistant}, we use the MPMQA \cite{mpmqa} dataset, comprising 250 pages of documents for 36 products. For \textbf{agriculture}, we use the AgriQA dataset~\cite{agriqa} with 250 queries on crop management and equipment operation. For \textbf{technical support}, we sample 250 queries from TechQA~\cite{techqa} containing real-world support questions. For \textbf{IoT security}, we use 250 queries on threat detection and best practices from a smart device security dataset~\cite{iotsecurity}. For automotive and smarthome assistant, we generate queries using our Context Generator; for other domains, we use existing queries and responses in the datasets directly. We measure response quality using G-Eval~\cite{geval} with an ensemble of two LLM judges (GPT-4o and Gemini-2.5-Flash) to mitigate self-preference bias~\cite{wataoka2025selfpreferencebias}—neither judge appears as a cloud model in our inference pipeline. We measure latency as Time to First Token (TTFT) and cost using OpenAI pricing (\$/1k queries).

\textbf{Hardware platforms.} 
We evaluate \name on four edge-hardware platforms spanning resource-constrained to high-performance options : Jetson Orin Nano (8GB RAM, 33 TOPS, 15 Watts, \$200), M1 Pro (16GB RAM, 11 TOPS, 45 Watts, \$1,000), M4 (32GB RAM, 38 TOPS, 65 Watts, \$700), and RTX A4500 (20GB VRAM, 186 TFLOPS, 200 Watts, \$1,300).


\textbf{Baselines.} We compare against RouteLLM~\cite{routellm}, a state-of-the-art model routing system, augmented with our RAG module for fair comparison. We evaluate three RouteLLM modes: RouteLLM-25, -50, and -75, with the number indicating percentage of queries sent to the cloud. We include GPT-4.1 as the cloud-only baseline and Oracle (exhaustive path selection) as the upper bound.

\textbf{\name configurations.} We evaluate two practical configurations: \name(Cost-First) minimizes operational cost, while \name(Latency-First) optimizes latency. Both use three edge models (SmolLM2-1.7B~\cite{smollm2smolgoesbig}, Llama3.2-3B~\cite{llama3}, Phi-4~\cite{phi4}) and three cloud options (GPT-4.1-nano, -mini, -full~\cite{openai}), combined with query processing (step-back prompting~\cite{stepback}), retrieval (basic RAG~\cite{basicrag}, HyDE~\cite{hyde}), and context processing (Corrective-RAG~\cite{correctiverag}, reranking~\cite{reranking}), yielding 200-300 distinct paths per domain.
\vspace{-1em}

\begin{table*}[t]
\centering
\small
\caption{Performance across hardware platforms on automotive and smart home domains. M4-class platforms emerge as the practical deployment target—achieving sub-second latencies (0.7s automotive, 2.3s smart home) with 74-82\% accuracy at \$2.2-2.4 per 1k queries. Lower-tier platforms (Orin) exhibit prohibitive latencies (12+ seconds), while higher-tier platforms (A4500) show no latency advantage despite high rated power consumption. \name maintains consistent accuracy (70-85\%) and minimal selection overhead (28-50ms) across all platforms.}
\label{tab:hardware_results}
\setlength{\tabcolsep}{4pt}
\begin{tabular}{lccccccc}
\toprule
\textbf{Hardware} & \textbf{Oracle} & \textbf{GPT-4.1} & \textbf{R-25} & \textbf{R-50} & \textbf{R-75} & \textbf{ECO-C} & \textbf{ECO-L} \\
\midrule
\multicolumn{8}{l}{\textit{Automotive Assistant}} \\
\hline
A4500 & 94\,/\,1.7\,/\,2.9 & 86\,/\,10.9\,/\,10.4 & 70\,/\,2.5\,/\,7.1 & 75\,/\,5.4\,/\,7.3 & 80\,/\,8.1\,/\,7.4 & 78\,/\,\textcolor{blue}{0.7}\,/\,0.8(41) & \textcolor{dkgreen}{82}\,/\,3.1\,/\,\textcolor{mauve}{0.7(37)} \\
M4 & 95\,/\,1.7\,/\,4.1 & 89\,/\,12.3\,/\,1.0 & 73\,/\,3.5\,/\,4.3 & 80\,/\,7.3\,/\,3.0 & \textcolor{dkgreen}{84}\,/\,9.9\,/\,2.2 & 82\,/\,\textcolor{blue}{2.4}\,/\,1.2(50) & 82\,/\,5.3\,/\,\textcolor{mauve}{0.7(33)} \\
M1Pro & 94\,/\,1.6\,/\,3.9 & 88\,/\,11.0\,/\,15.2 & 72\,/\,2.2\,/\,14.8 & 78\,/\,5.5\,/\,14.1 & 82\,/\,8.3\,/\,13.4 & 79\,/\,\textcolor{blue}{1.6}\,/\,\textcolor{mauve}{1.1(48)} & \textcolor{dkgreen}{83}\,/\,5.6\,/\,1.6(28) \\
Orin & 96\,/\,1.8\,/\,14.7 & 98\,/\,12.9\,/\,41.7 & 66\,/\,\textcolor{blue}{0.3}\,/\,54.1 & 68\,/\,0.7\,/\,54.1 & 70\,/\,1.9\,/\,53.2 & \textcolor{dkgreen}{85}\,/\,5.2\,/\,16.6(45) & \textcolor{dkgreen}{85}\,/\,4.2\,/\,\textcolor{mauve}{12.6(37)} \\
\midrule
\multicolumn{8}{l}{\textit{Smart Home Assistant}} \\
\hline
A4500 & 90\,/\,1.4\,/\,2.5 & 75\,/\,4.8\,/\,0.8 & 58\,/\,\textcolor{blue}{1.1}\,/\,\textcolor{mauve}{0.8} & 63\,/\,1.9\,/\,0.8 & 71\,/\,3.3\,/\,0.9 & 74\,/\,2.5\,/\,2.2(39) & \textcolor{dkgreen}{75}\,/\,2.6\,/\,2.0(37) \\
M4 & 91\,/\,1.9\,/\,4.6 & 73\,/\,8.8\,/\,24.8 & 54\,/\,\textcolor{blue}{2.0}\,/\,22.6 & 59\,/\,3.4\,/\,22.6 & 66\,/\,5.9\,/\,22.0 & \textcolor{dkgreen}{74}\,/\,2.2\,/\,4.4(36) & 73\,/\,3.3\,/\,\textcolor{mauve}{2.3(37)} \\
M1Pro & 90\,/\,1.7\,/\,3.2 & 75\,/\,4.8\,/\,0.8 & 56\,/\,1.6\,/\,6.8 & 61\,/\,2.7\,/\,6.8 & 68\,/\,4.6\,/\,6.8 & \textcolor{dkgreen}{74}\,/\,\textcolor{blue}{0.7}\,/\,3.1(36) & 73\,/\,1.5\,/\,\textcolor{mauve}{1.5(35)} \\
Orin & 90\,/\,1.8\,/\,9.3 & 75\,/\,4.8\,/\,0.9 & 58\,/\,\textcolor{blue}{1.1}\,/\,1.1 & 64\,/\,1.9\,/\,1.2 & 71\,/\,3.3\,/\,\textcolor{mauve}{1.0} & 70\,/\,3.0\,/\,3.9(34) & \textcolor{dkgreen}{73}\,/\,3.8\,/\,6.1(31) \\
\bottomrule
\multicolumn{8}{l}{\footnotesize{Format: Accuracy (\%) / Cost (\$/1k queries) / Latency (seconds). Parentheses: ECO-LLM selection overhead (ms).}} \\
\multicolumn{8}{l}{\footnotesize{R-25/50/75: RouteLLM with 25\%/50\%/75\% cloud routing (latency includes routing overhead). ECO-C/L: \name(Cost/Latency-First).}} \\
\multicolumn{8}{l}{\footnotesize{\textcolor{dkgreen}{Green}: Best accuracy; \textcolor{blue}{Blue}: Best cost; \textcolor{mauve}{Mauve}: Best latency among RouteLLM and \name variants.}} \\
\end{tabular}
\vspace{-1em}
\end{table*}

\subsection{Performance Across Edge Hardware}

We evaluate cross-platform performance on the automotive and smart home domains, since they have different pre-processing requirements—automotive queries require extensive retrieval while smart home queries benefit minimally from pre-processing. Our results shown in Table ~\ref{tab:hardware_results} identifies M4-class platforms as a particularly cable edge device for deplopying responsive edge AI assistants given it  integrated NPUs, 30-40 TOPS compute, and 20-32GB unified RAM within a 65W power envelope. As shown in Table~\ref{tab:hardware_results}, the M4 achieves responsive latencies with \name(Latency-First) delivering 0.7s for automotive and 2.3s for smart home, while \name(Cost-First) operates at \$2.24-2.41 per 1k queries with 74-82\% accuracy.

In comparison, the Jetson Orin Nano (8GB, 33 TOPS) has latencies exceeding 12s for automotive queries, making it unsuitable for interactive applications despite achieving higher accuracy through aggressive cloud routing. The M1 Pro (16GB, 11 TOPS), despite costing 43\% more than the M4, provides lower compute (11 vs 38 TOPS) and has no meaningful advantage—\name configurations achieve comparable accuracy and latency while the platform's higher cost reduces deployment flexibility.

Conversely, additional resources beyond M4-class yield diminishing returns. The RTX A4500 (186 TFLOPS, 200W) shows no latency advantage despite substantially higher power consumption—\name(Latency-First) achieves 0.7s for automotive on both platforms. Furthermore, the A4500's power and cooling requirements limit deployment to infrastructure rich scenarios, whereas M4-class platforms enable deployment in vehicles, homes, and mobile contexts.

Across all platforms, \name maintains consistent accuracy (73-85\%) and minimal selection overhead (33-50ms), confirming that component selection strategies transfer across hardware. We focus our subsequent evaluation on the M4. 

\begin{table*}[t]
\centering
\small
\caption{Performance across five domains on M4 hardware. All baselines and GPT-4.1 use the best-average preprocessing configuration from emulation for fair comparison. ECO-LLM achieves lowest latency in all domains and highest accuracy in three, with consistent 73-87\% accuracy compared to RouteLLM's 54-85\% variance. Averaging across domains: ECO-LLM reduces costs by 61\% (\$2.74 vs \$7.08) and latency by 6x (1.7s vs 10.6s) compared to RouteLLM-75 while maintaining comparable accuracy (79-81\% vs 80\%). Oracle outperforms GPT-4.1 (94\% vs 85\% average accuracy), demonstrating that joint optimization yields superior results even compared to the best cloud model with optimized preprocessing.}
\label{tab:domain_results}
\setlength{\tabcolsep}{4pt}
\begin{tabular}{lccccccc}
\toprule
\textbf{Domain} & \textbf{Oracle} & \textbf{GPT-4.1} & \textbf{R-25} & \textbf{R-50} & \textbf{R-75} & \textbf{ECO-C} & \textbf{ECO-L} \\
\midrule
Agriculture & 96\,/\,0.6\,/\,3.1 & 87\,/\,5.8\,/\,1.0 & 80\,/\,1.1\,/\,1.6 & 82\,/\,2.3\,/\,1.5 & \textcolor{dkgreen}{83}\,/\,3.6\,/\,1.3 & 79\,/\,\textcolor{blue}{0.2}\,/\,1.4(32) & 77\,/\,0.3\,/\,\textcolor{mauve}{1.2(43)} \\
TechQA & 95\,/\,6.5\,/\,11.5 & 87\,/\,15.5\,/\,18.0 & 66\,/\,4.6\,/\,21.9 & 74\,/\,8.6\,/\,21.5 & 80\,/\,11.8\,/\,21.0 & \textcolor{dkgreen}{84}\,/\,4.1\,/\,5.3(64) & 81\,/\,\textcolor{blue}{3.7}\,/\,\textcolor{mauve}{1.3(48)} \\
IoT Security & 94\,/\,1.2\,/\,3.4 & 90\,/\,7.1\,/\,6.3 & 82\,/\,\textcolor{blue}{1.8}\,/\,6.6 & 85\,/\,3.3\,/\,6.6 & 85\,/\,4.2\,/\,6.6 & \textcolor{dkgreen}{87}\,/\,4.8\,/\,5.7(44) & 84\,/\,4.4\,/\,\textcolor{mauve}{3.1(40)} \\
Automotive & 95\,/\,1.7\,/\,4.1 & 89\,/\,12.3\,/\,1.0 & 73\,/\,3.5\,/\,4.3 & 80\,/\,7.3\,/\,3.0 & \textcolor{dkgreen}{84}\,/\,9.9\,/\,2.2 & 82\,/\,\textcolor{blue}{2.4}\,/\,1.2(50) & 82\,/\,5.3\,/\,\textcolor{mauve}{0.7(33)} \\
Smart Home & 91\,/\,1.9\,/\,4.6 & 73\,/\,8.8\,/\,24.8 & 54\,/\,\textcolor{blue}{2.0}\,/\,22.6 & 59\,/\,3.4\,/\,22.6 & 66\,/\,5.9\,/\,22.0 & \textcolor{dkgreen}{74}\,/\,2.2\,/\,4.4(36) & 73\,/\,3.3\,/\,\textcolor{mauve}{2.3(37)} \\
\bottomrule
\multicolumn{8}{l}{\footnotesize{Format: Accuracy (\%) / Cost (\$/1k queries) / Latency (seconds). Parentheses: ECO-LLM selection overhead (ms).}} \\
\multicolumn{8}{l}{\footnotesize{R-25/50/75: RouteLLM with 25\%/50\%/75\% cloud routing (latency includes routing overhead). ECO-C/L: \name(Cost/Latency-First).}} \\
\multicolumn{8}{l}{\footnotesize{\textcolor{dkgreen}{Green}: Best accuracy; \textcolor{blue}{Blue}: Best cost; \textcolor{mauve}{Mauve}: Best latency among RouteLLM and \name variants.}} \\
\end{tabular}
\vspace{-1em}
\end{table*}

\subsection{Generalization Across Domains}

We evaluate \name's ability to generalize across five diverse domains on the M4 hardware. Table~\ref{tab:domain_results} shows that \name achieves lowest latency across all five domains and highest accuracy in three, using the same system configuration.

\name(Latency-First) reduces response times substantially across domains: 1.3s for TechQA versus RouteLLM-75's 21s, 2.3s for Smart Home versus 22s, and 0.7s for Automotive versus 2.2s. \name(Cost-First) achieves highest accuracy in Smart Home (74\% vs RouteLLM-75's 66\%), TechQA (84\% vs 80\%), and IoT Security (87\% vs 85\%). In Automotive and Agriculture, \name configurations reach 77-82\% accuracy compared to RouteLLM-75's 83-84\%, while reducing costs by 4-18×—a tradeoff that may suit latency or cost-sensitive deployments.

In comparison, RouteLLM shows higher variance in accuracy across domains, ranging from 54\% (Smart Home, R-25) to 85\% (IoT, R-50/R-75). In Smart Home, RouteLLM-75 achieves only 66\% accuracy with 22s latency; in TechQA, all RouteLLM configurations exhibit 20+ second latencies regardless of cloud routing percentage. These patterns suggest that model selection alone is insufficient when queries benefit from different preprocessing and retrieval strategies, which \name's joint optimization addresses.

\name provides predictable configuration options across all domains: Cost-First delivers cost-efficiency while Latency-First provides responsiveness, while keeping accuracy within 73-87\% compared to RouteLLM's 54-85\%. Practitioners can select based on deployment priorities without domain-specific tuning. The emulator discovers domain-appropriate component patterns automatically—the same training procedure yields different critical component configurations for each domain. Furthermore, the path selection overhead remains minimal (32-64ms).

\begin{table*}[t]
\centering
\small
\caption{Ablation study validating ECO-LLM components. Static policies sacrifice secondary metrics (Cost-First: 10.2s latency; Latency-First: \$6.1 cost). CCA alone degrades accuracy; DSQE's learned encoding is necessary to optimize both dimensions simultaneously.}
\label{tab:ablation_results}
\setlength{\tabcolsep}{4pt}
\begin{tabular}{lcccccc}
\toprule
\multirow{2}{*}{\textbf{Domain}} & \multicolumn{3}{c}{\textbf{Cost-First Optimization}} & \multicolumn{3}{c}{\textbf{Latency-First Optimization}} \\
\cmidrule(lr){2-4} \cmidrule(lr){5-7}
 & Static & CCA Only & \textbf{ECO-C} & Static & CCA Only & \textbf{ECO-L} \\
\midrule
AgriQA & 81.3\,/\,0.3\,/\,7.3 & 76.0\,/\,0.5\,/\,5.1 & \textbf{79.0\,/\,0.2\,/\,1.4} & 84.5\,/\,0.3\,/\,0.7 & 76.2\,/\,0.7\,/\,4.3 & \textbf{77.0\,/\,0.3\,/\,1.2} \\
IoT Security & 88.8\,/\,2.2\,/\,8.1 & 85.4\,/\,4.0\,/\,5.3 & \textbf{87.0\,/\,4.8\,/\,5.7} & 85.2\,/\,5.7\,/\,2.9 & 83.5\,/\,4.8\,/\,4.4 & \textbf{84.0\,/\,4.4\,/\,3.1} \\
SmartCars & 82.7\,/\,0.7\,/\,17.2 & 78.6\,/\,1.7\,/\,4.7 & \textbf{82.0\,/\,2.4\,/\,1.2} & 88.6\,/\,12.3\,/\,1.0 & 78.2\,/\,4.1\,/\,2.6 & \textbf{82.0\,/\,5.3\,/\,0.7} \\
SmartHome & 70.3\,/\,0.3\,/\,0.7 & 72.3\,/\,1.1\,/\,3.6 & \textbf{74.0\,/\,2.2\,/\,4.4} & 75.1\,/\,4.8\,/\,0.9 & 74.1\,/\,2.5\,/\,2.8 & \textbf{73.0\,/\,3.3\,/\,2.3} \\
TechQA & 85.2\,/\,4.1\,/\,17.8 & 82.5\,/\,7.6\,/\,14.9 & \textbf{84.0\,/\,4.1\,/\,5.3} & 86.8\,/\,7.4\,/\,0.9 & 82.5\,/\,10.5\,/\,9.7 & \textbf{81.0\,/\,3.7\,/\,1.3} \\
\midrule
\textbf{Average} & 81.7\,/\,1.5\,/\,10.2 & 79.0\,/\,3.0\,/\,6.7 & \textbf{81.2\,/\,2.7\,/\,3.6} & 84.0\,/\,6.1\,/\,1.3 & 78.9\,/\,4.5\,/\,4.8 & \textbf{79.4\,/\,3.4\,/\,1.7} \\
\bottomrule
\multicolumn{7}{l}{\footnotesize{Format: Accuracy (\%) / Cost (\$/1k queries) / Latency (seconds).}} \\
\multicolumn{7}{l}{\footnotesize{Static: Fixed path for all queries; CCA Only: CCA with semantic matching (no DSQE).}} \\
\end{tabular}
\vspace{-1em}
\end{table*}

\subsection{Contributions of \name Components}

We conduct systematic ablation studies to evaluate the necessity of \name's adaptive runtime selection and the role of its components—Critical Component Analysis (CCA) and Domain-Specific Query Encoding (DSQE). We compare three configurations across all five domains on M4 hardware:

\textbf{Config 1 - Static Policy (No CCA, No DSQE):} Represents the traditional deployment approach where practitioners select a single ``best average'' query resolution path based on training data performance, then apply it uniformly to all queries. The path is chosen by (1) filtering paths within an accuracy margin of the best-performing configuration, then (2) selecting the lowest-cost or lowest-latency option based on deployment priority. This baseline requires no runtime selection logic.

\textbf{Config 2 - CCA Only (No DSQE):} Introduces runtime per-query path selection using full CCA to identify critical components, but employs direct 1-nearest-neighbor matching with sentence transformer embeddings instead of DSQE's learned projection. This configuration isolates DSQE's contribution while demonstrating basic adaptive selection (selection overhead: 20-30ms).

\textbf{Config 3 - Full \name (CCA + DSQE):} The full \name system with both CCA and DSQE's learned query encoding, as evaluated earlier (selection overhead: 30-50ms).

As shown in Table~\ref{tab:ablation_results}, Static (Config 1) Cost-First averages \$1.5 per 1k queries across all the datasets but requires 10.2s latency—unsuitable for interactive applications. Conversely, Static (Config 1) Latency-First achieves 1.3s latency but costs \$6.1, reflecting that a single fixed path cannot simultaneously serve queries with different component requirements.
Introducing adaptive per-query selection addresses this limitation, but the encoding mechanism proves critical. The CCA-only configuration (Config 2)—which identifies critical components correctly but relies on 1-nearest-neighbor matching with semantic embeddings—actually degrades accuracy compared to static policies: 79.0\% versus 81.7\% for Cost-First, 78.9\% versus 84.0\% for Latency-First. This occurs because semantic similarity does not reliably indicate component requirements; queries with similar surface forms may need different pre-processing and retrieval strategies, as noted in Section 3.3.
DSQE's learned projection resolves this mismatch. Full ECO-LLM restores accuracy to 81.2\% and 79.4\% respectively while substantially improving secondary metrics—Cost-First latency drops from 10.2s to 3.6s (3×), Latency-First cost drops from \$6.1 to \$3.4 (1.8×). By learning domain-specific patterns that map queries to appropriate component sets, DSQE enables effective per-query adaptation that static policies or naive semantic matching can't achieve.
\vspace{-1em}

\begin{figure}[t]
\centering
    \subfigure[Latency - Violation]{
        \label{fig:latency_slo_all}
        \includegraphics[width=0.22\textwidth]{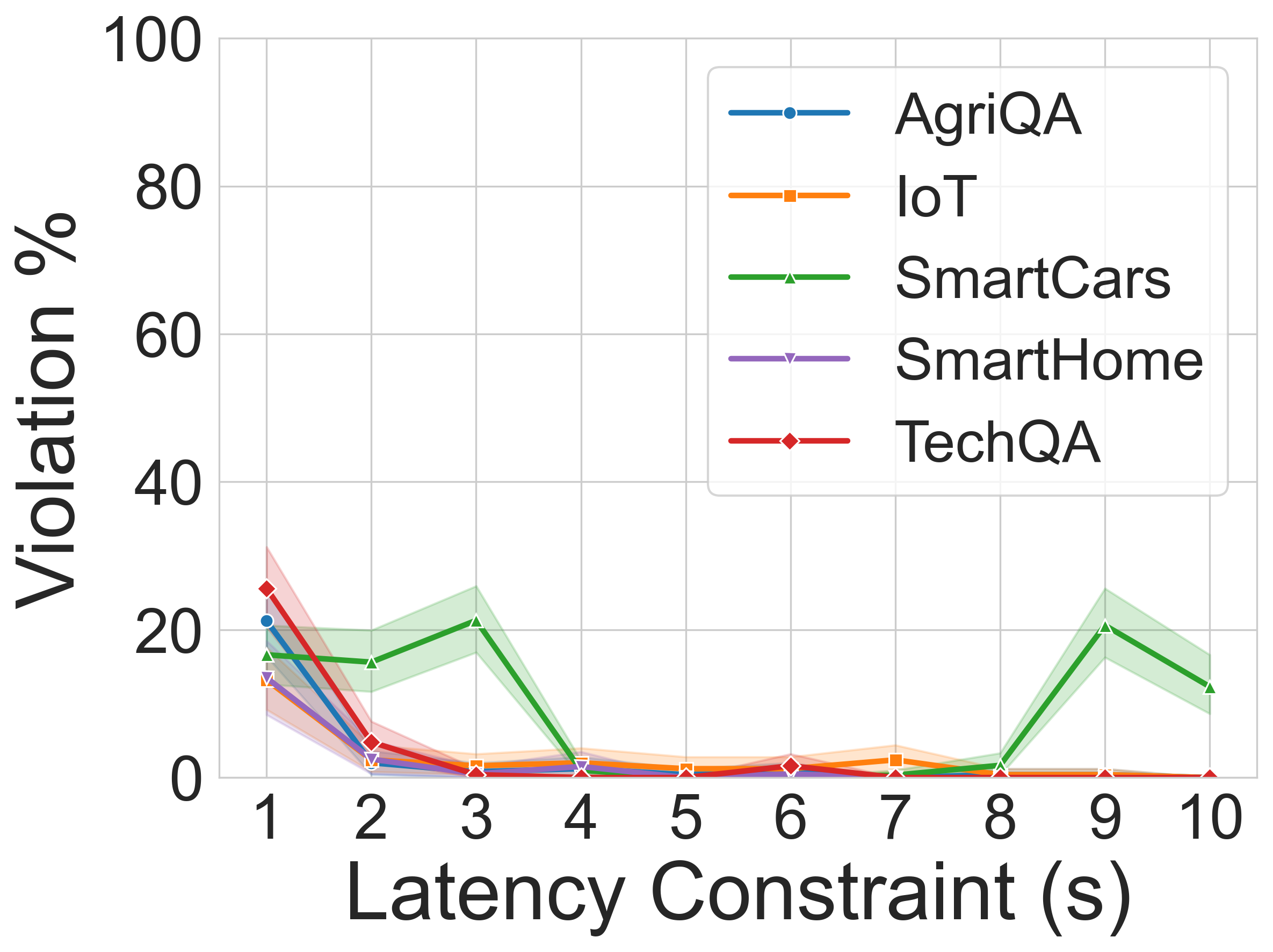}}
    \hfill
    \subfigure[Latency - Accuracy]{
        \label{fig:latency_performance_all}
        \includegraphics[width=0.22\textwidth]{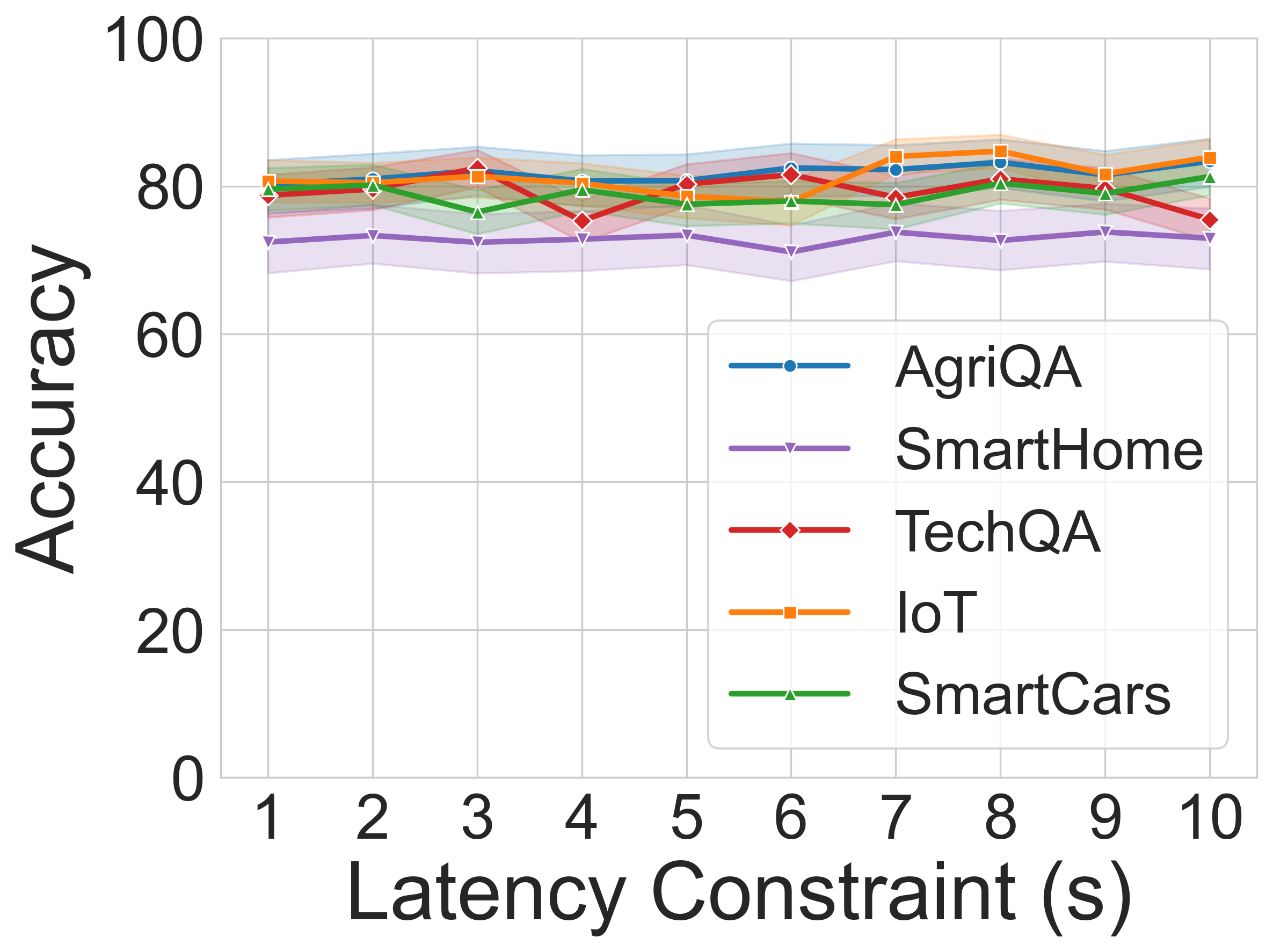}}
    
    \vspace{2mm}
    
    \subfigure[Cost - Violation]{
        \label{fig:cost_slo_all}
        \includegraphics[width=0.22\textwidth]{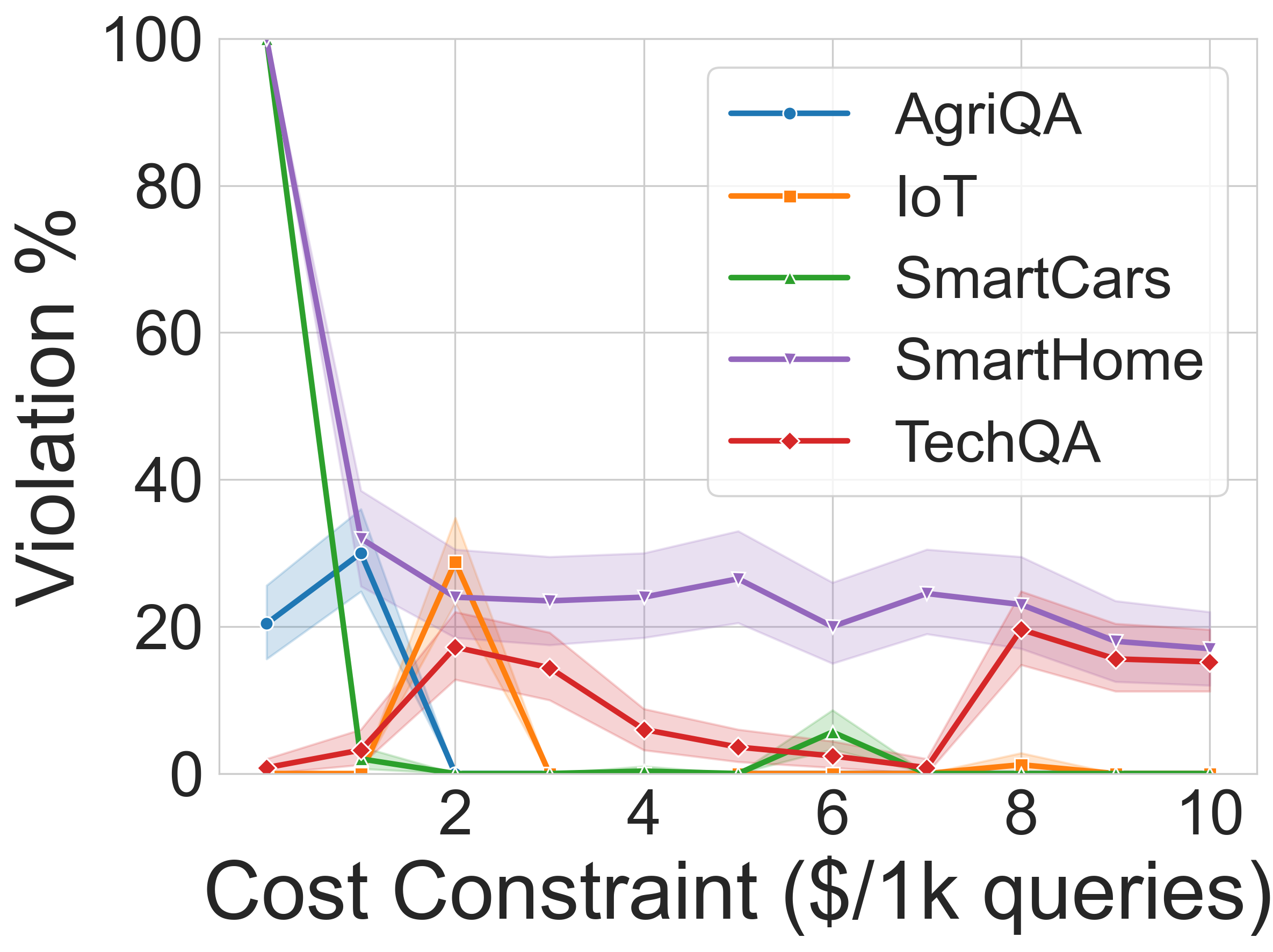}}
    \hfill
    \subfigure[Cost - Accuracy]{
        \label{fig:cost_performance_all}
        \includegraphics[width=0.22\textwidth]{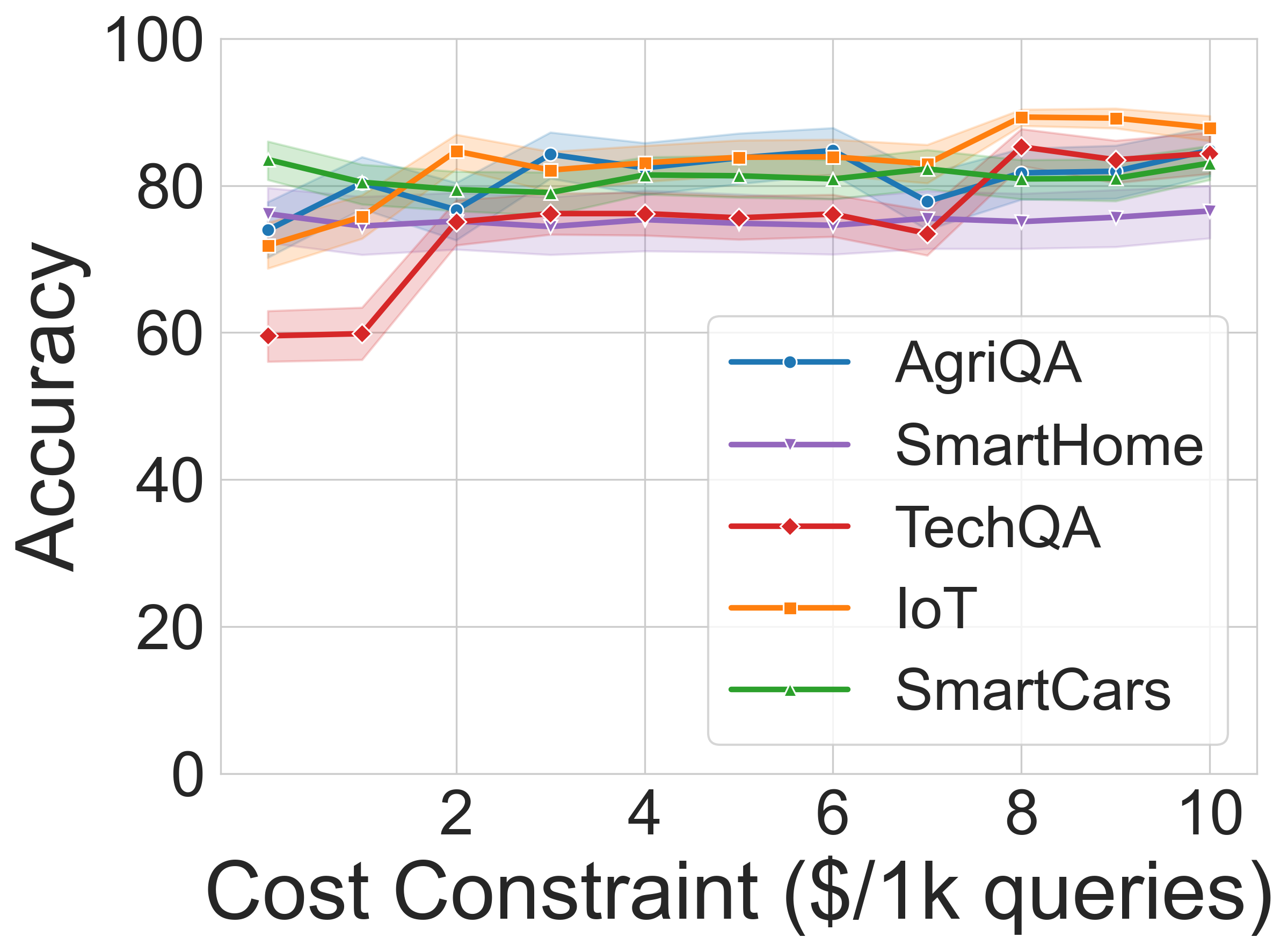}}

\caption{SLO attainment across four domains. Figures (a) and (b) show latency SLO results for \name(Latency-First). Figures (c) and (d) show cost SLO results for \name(Cost-First). Shaded regions represent 99\% confidence intervals.}
\label{fig:slo_all_domains}
\vspace{-1em}
\end{figure}

\subsection{SLO Attainment}
We evaluate \name's ability to meet user-defined Service Level Objectives (SLO) across all five domains on the M4 hardware. Figure~\ref{fig:slo_all_domains} shows violation rates and accuracy as constraints vary from strict to relaxed—cost from \$0-10 per 1k queries, latency from 1-10 seconds.
\name achieves near-zero violations when constraints are achievable. For latency SLOs (Figures~\ref{fig:latency_slo_all},\ref{fig:latency_performance_all}), AgriQA, IoT, SmartHome, and TechQA drop below 5\% violation rate at 2-second constraints and approach zero beyond 4 seconds. For cost SLOs (Figures\ref{fig:cost_slo_all},~\ref{fig:cost_performance_all}), AgriQA, IoT, and SmartCars reach near-zero violations at \$3-4 per 1k queries.
Some domain-constraint combinations show persistent violations—SmartHome shows 15-25\% cost violations even at \$10, and TechQA shows 15-20\% violations at higher cost constraints. These reflect \name's design choice to prioritize response quality over strict SLO compliance. As described in Section~\ref{sec:orchestration_design}, the runtime first identifies critical components for accuracy, then selects the lowest-cost or lowest-latency path with those components. When critical components inherently exceed the constraint, \name exceeds the SLO rather than using cheaper alternatives that reduce accuracy.
We believe that this design is appropriate for domain-specific assistants where incorrect responses provide no practical value regardless of cost or speed. The accuracy plots (Figures~\ref{fig:latency_performance_all},~\ref{fig:cost_performance_all}) validate this approach: accuracy remains stable at 73-85\% across all constraint levels. Strict SLO compliance would instead show accuracy degradation as constraints tighten. Furthermore, persistent violations are a signal to practitioners that their constraints are too restrictive for the domain's query distribution—information that silent quality degradation would obscure.
\vspace{-1em}


\begin{table}[t]
\centering
\small
\caption{Accuracy delta from full (100\%) exploration across varying exploration budgets. SBA maintains near-equivalent accuracy with reduced exploration—at 70\% (B=10), accuracy stays within 1.5 points; at 35\% (B=5), most domains remain within 3 points. \textcolor{dkgreen}{Green} indicates constrained exploration outperformed full exploration.}
\label{tab:budget_impact}
\setlength{\tabcolsep}{3pt}
\begin{tabular}{@{}lrrrrrr@{}}
\toprule
\multirow{2}{*}{\textbf{Domain}} & \multicolumn{3}{c}{\textbf{Cost-First}} & \multicolumn{3}{c}{\textbf{Latency-First}} \\
\cmidrule(lr){2-4} \cmidrule(lr){5-7}
 & \textbf{14\%} & \textbf{35\%} & \textbf{70\%} & \textbf{14\%} & \textbf{35\%} & \textbf{70\%} \\
\midrule
AgriQA & $-$0.7 & $-$1.2 & \textcolor{dkgreen}{$+$3.3} & $-$3.6 & \textcolor{dkgreen}{$+$0.3} & $-$1.5 \\
IoT Security & $-$8.1 & $-$7.2 & \textcolor{dkgreen}{$+$0.1} & \textcolor{dkgreen}{$+$2.3} & \textcolor{dkgreen}{$+$0.5} & \textcolor{dkgreen}{$+$0.4} \\
Automotive & $-$2.0 & \textcolor{dkgreen}{$+$0.6} & \textcolor{dkgreen}{$+$0.5} & $-$7.2 & $-$0.1 & $-$0.6 \\
Smart Home & — & $-$0.1 & \textcolor{dkgreen}{$+$0.8} & $-$4.5 & $-$4.5 & $-$0.04 \\
TechQA & $-$2.5 & $-$2.4 & $-$0.1 & \textcolor{dkgreen}{$+$2.9} & \textcolor{dkgreen}{$+$3.5} & \textcolor{dkgreen}{$+$1.2} \\
\bottomrule
\end{tabular}

\vspace{0.5em}
\raggedright
\footnotesize{14\% (B=2), 35\% (B=5), 70\% (B=10): percentage of queries explored.}\\
\footnotesize{Values: accuracy delta from 100\% exploration (percentage points).}
\vspace{-1em}
\end{table}

\subsection{Exploration Budget Efficiency}

We evaluate whether SBA's stratified sampling provides sufficient exploration when computational resources constrain exhaustive evaluation. Table~\ref{tab:budget_impact} shows accuracy differences relative to full exploration across three reduced budget levels—14\% (B=2), 35\% (B=5), and 70\% (B=10) of queries explored.

At 70\% exploration, accuracy remains within 1.5 percentage points of full exploration across all domains and optimization strategies, confirming that reduced exploration introduces minimal degradation. At 35\% exploration, most domain-strategy combinations stay within 3 percentage points—AgriQA, Automotive, and TechQA show differences of 2.4\% or less for both strategies, while IoT Security (Cost-First) and Smart Home (Latency-First) exhibit larger gaps of 7.2\% and 4.5\% respectively. These outliers suggest domains with more heterogeneous query distributions benefit from additional exploration.

The results exhibit non-monotonic patterns—several configurations show positive deltas where constrained exploration outperforms full exploration. TechQA Latency-First achieves 3.5\% higher accuracy at 35\% exploration than at 100\%, and IoT Security Latency-First shows consistent improvements across all reduced budgets. This suggests that exhaustive exploration can discover paths that marginally improve training performance but generalize poorly, whereas SBA's clustering-based selection focuses on representative queries and yields more robust strategies.
Overall, SBA provides effective subsampling when exploration budget is constrained. Practitioners prioritizing reliability can use 70\% exploration for near-equivalent performance, while those with tighter constraints can use 35\% with the understanding that complex domains may see modest accuracy reductions.

\vspace{-1em}

\section{Discussion}
\vspace{-1em}

\modified{\name follows a domain-specific deployment model where the system trains once per domain and deploys a runtime optimized for that domain's characteristics. We discuss several considerations for deployment and future directions.}

\modified{\textbf{Query Distribution and Concept Drift.}
DSQE handles unseen queries within the training distribution but performance degrades for queries that significantly deviate from training patterns. This stems from DSQE's reliance on learned prototypes representing component sets observed during training. When query distributions shift over time—such as new vehicle features or device types not in original documentation—the critical component patterns may change, requiring retraining with updated documentation and regenerated queries. The performance gap between seen and unseen queries suggests potential for improved generalization through better query feature engineering and more sophisticated metric prediction models.}

\modified{\textbf{Hardware Portability and Device-Specific Profiling.}
The \emulator profiles component performance on a single hardware configuration, producing platform-specific latency and cost metrics. Deploying to different edge devices requires profiling on target hardware, using virtual environments with configurable resources, or employing latency prediction models like nn-Meter to translate metrics across devices. Incorporating hardware specifications as runtime parameters and developing cross-device performance transfer methods would enhance \name's applicability across diverse platforms.}

\modified{\textbf{Continuous Adaptation and Online Learning.}
The current implementation uses one-shot training where both CCA and DSQE train offline and remain fixed during deployment. CCA identifies critical component sets through ablation studies on training queries, and DSQE learns to predict these sets for new queries. In production, actual query distributions may differ from generated training queries, and performance requirements may evolve. While the runtime adapts path selection to changing SLO constraints, the offline training phase (CCA and DSQE) cannot adjust to new query patterns without complete retraining. Developing online learning mechanisms that continuously update both modules based on real-world feedback would enable adaptation to evolving workloads. This requires tracking low-confidence predictions, identifying emerging patterns, re-evaluating component criticality for new query types, and incrementally updating prototype representations.}

\vspace{-1em}

\section{Conclusion}
\vspace{-1em}

We present \name, a framework for building efficient domain-specific edge AI assistants that optimizes the complex interplay between query resolution components. \name enables systematic evaluation of resolution paths and intelligent runtime adaptation while requiring minimal domain expertise from users. Our evaluation demonstrates that \name achieves comparable performance to existing solutions while operating at the Pareto-optimal front for both cost and latency, as well as meeting user-defined SLOs across different domains. This combination of systematic optimization and runtime adaptability establishes \name as a comprehensive solution for deploying and managing edge AI assistants in real-world applications.

\bibliographystyle{plain}
\bibliography{references}

\end{document}